\documentstyle[12pt,a4,epsfig]{article}

\newcommand{\aprle}          {\buildrel < \over {_{\sim}}}
\newcommand{\aprge}          {\buildrel > \over {_{\sim}}} 

\newcommand{\joudef}[2]{\newcommand #1{\jou{\ignorespaces #2}}}
\joudef{\prd}    { Phys.\ Rev.\ D}

\begin{document}
\newcommand{\lsim}{\lower .5ex\hbox{$\buildrel < \over {\sim}$}}
\newcommand{\gsim}{\lower .5ex\hbox{$\buildrel > \over {\sim}$}}

\title{A parameterisation of single and multiple muons in the deep water or ice}
\author{Y. Becherini, A. Margiotta, M. Sioli and M. Spurio \\
\small Department of Physics of the University of Bologna and INFN Sezione di 
Bologna, \\ 
\small Viale Berti Pichat 6/2, I-40127 Bologna, Italy }

\maketitle

\begin{abstract}
Atmospheric muons play an important role in underwater/ice neutrino detectors. In this paper, a parameterisation of the flux of single and multiple muon events, their lateral distribution  and of their energy spectrum is presented. The kinematics parameters were modelled starting from a full Monte Carlo simulation of the interaction of primary cosmic rays with atmospheric nuclei; secondary muons reaching the sea level were propagated in the deep water.
The parametric formulas are valid for a vertical depth of $1.5\div 5\ km \:w.e.$ and up to $85^\circ$ for the zenith angle, and  can be used as input for a fast simulation of atmospheric muons in underwater/ice detectors.
\end{abstract}

\section{Introduction \label{sec:intro} }

The realization of a $km^3$ scale detector for astrophysical neutrinos  is today considered one of the most important challenges of the next decade for high energy physics and astrophysics \cite{learned}. Up to now, reduced scale detectors and prototypes have demonstrated the feasibility of the Cherenkov detection technique, while larger projects are under way \cite{piattelli14}.

The study of muon bundles in deep underground experiments was originally motivated because it could provide some clues on primary cosmic ray flux and composition in the PeV region \cite{gasta}. For neutrino astronomy experiments, where neutrino-induced muons are discriminated by selecting upward going tracks, atmospheric muons can represent a major background source. Downward-going muons can incorrectly be reconstructed as upward-going particles and mimic high energy neutrino interactions; muons in bundles seem to be particularly dangerous.
Simulations \cite{comparison} show that a large fraction (at least 40\%) of wrongly reconstructed upward going events are induced by muons in bundles. These muons are expected to arrive almost at the same time in the plane perpendicular to the shower axis. Experimental data \cite{macroarival} show that the arrival time spread of muons in bundles is smaller than few $ns$. 

Direct measurements of the muon flux at sea level exist, as well as parametric formulas \cite{miyake}.  On the contrary,  atmospheric muon flux in the deep water is  only roughly known experimentally \cite{grieder}. Some  parameterisations for the energy spectrum of underwater single muons versus the vertical depth $h$ and the zenith angle $\theta$ are available in literature \cite{okada,bugaev,klimu}. These calculations are based on the parametric description of the muon flux at the sea level, and use semi-analytic methods to solve the muon transport equation in a homogeneous medium. In any case, all authors assume a flux of single muons only. 

In this paper, for the first time, parametric formulas are given to calculate the flux of muon bundles, taking into account the muon multiplicity and the muon energy spectrum in a bundle, as a function of the distance from the shower axis. 
The range of validity extends from 1.5 km down to 5.0 km of water vertical depth, and from $0^\circ $ up to $85^\circ $ for the zenith angle. 
A full Monte Carlo simulation (section \ref{sec:MC}) was used to evaluate all the kinematics parameters of muons arriving at seven depths. 
The muon energy spectrum depends on the water/ice vertical depth $h$, on the zenith angle $\theta$, on the bundle multiplicity $m$ and, for each muon in the bundle, on the distance of the muon with respect to the shower axis. 
The  parameterisation provides a self-consistent generator which can be used as input by underwater/ice experiments for the simulation of the atmospheric muon flux.
In the equations reported in this paper,  the following units are used: depth $h$ in $km \:w.e. = 10^6 \ kg\ m^{-2}$; zenith angle $\theta$ in $radians$; muon energies $E_\mu$ in $TeV$;
radial distances $R$ in $m$.

The plan of the work paper is as follows. In sec. \ref{sec:flux} the expression to evaluate the flux of muon bundles of any multiplicity is presented.
The depth-intensity relation for the vertical direction, as well as the dependence of the flux with the zenith angle compared with experimental data, is discussed in sec. \ref{sec:intensity}.
Since the largest fraction of events are single muons, in sec. \ref{sec:spectrum} the differential energy spectrum for single muon events as a function of the vertical depth and zenith angle is obtained. 
For muons arriving at the detector in bundles, the energy distribution depends also on the muon multiplicity and on the distance of each muon from the shower axis. This effect is studied in sec. \ref{sec:radial}, as well as the  parameterisation of the muon lateral spread function.
In sec. \ref{sec:multi}, the differential energy spectrum of muons in bundles is  parameterised in terms of all the four variables (muon multiplicity, vertical depth, zenith angle and radial distance). 
The comparison of expected results with the single and double muon average energy measured at different depths with the MACRO underground experiment is reported in sec. \ref{sec:macrotrd}.

\section{The Monte Carlo simulation \label{sec:MC} }

The  parameterisation of the multiple muon flux and energy spectra presented in this work relies on a full Monte Carlo simulation of primary Cosmic Rays (CR) interactions and shower propagation in the atmosphere (the HEMAS code, sec. 2.1). The adopted primary CR flux and composition model is described in sec. 2.2.
To optimise the showers production, the primary CRs have been divided in five mass groups, six primary energy intervals and two zenith angle regions. Each of these 60 files corresponds to a specific livetime for a $1\ km^2$ detector (the livetime spans from few hours for the low-energy bins to few years for the high energy ones).  The final results have been normalized to the same livetime. 

The muons from the decay of secondary mesons reaching the sea level are then propagated down to 5 km of water using the MUSIC code (sec. 2.3).
All the information concerning muons reaching the seven depths from $2.0\ km \:w.e.$ down to $5.0\ km \:w.e.$, in steps of $0.5\ km \:w.e.$, and of their primary CR parents are kept in ROOT files. The energy spectrum of muons depends, a part on the vertical depth $h$, on the zenith angle $\theta$, on the muon multiplicity in the shower $m$ and on the distance of the muon from the shower axis $R$. 

\subsection{The HEMAS code}

For the primary CR interaction and shower propagation in atmosphere, the HEMAS  (Hadronic, Electromagnetic and Muonic components in Air Showers) code is used \cite{hemas}.

HEMAS takes into account all the main physical
processes  occurring in the atmosphere: it computes the first interaction 
point of primary CR on the basis of the input cross sections, 
propagates electromagnetic and hadronic components of the showers considering 
the actual mean free path of particles, takes into account the deflection
of charged particles by the geomagnetic field. 
Hadronic interactions in the atmosphere are handled with the hadronic interaction code DPMJET \cite{dpmjet}. This is a model based on the two component Dual Parton Model (the hard and soft components). It also contains a detailed description of nuclear interactions. The number of nucleon-nucleon interactions is evaluated from the Glauber formalism. 
The mean free path of CR in the atmosphere is related to the inelastic cross sections 
of primary cosmic rays. For primary protons, we have
$\lambda_{p-air}(g/cm^{2})=\frac{2.4\times10^{4}}{\sigma_{p-air}(mb)}$ and similar relations hold for other primaries or secondary produced hadrons.

In this work the last version of the code (called HEMAS-DPM \cite{hemasdpm}) has been used. It includes the implementation of the Earth curvature, allowing to perform correct calculations at large zenith angles. Small angle approximations are used in the particle transport along the atmosphere, and the reliability of the code is restricted to secondary particles with energies $E\aprge$ 500 GeV.
Muons with energies $E < 500\ GeV$ (at sea level) are not followed through water; in any case, their contribution for depths larger than $1.5\ km \:w.e.$ is very small, and completely negligible at $2.0\ km \:w.e.$

The parameterisations are derived from histograms starting at $2.0\ km \:w.e.$ Nevertheless, the validity range of the results can be extended down to $\sim 1.5$ $km \:w.e.$, since no particular physical processes are expected to dominate at these depths, both for the hadro-production in the atmosphere and for the muon transport in water/ice.

The so called {\it prompt muons}, the component of the secondary cosmic ray flux originateds from the decay of charmed mesons and other short-lived particles produced in the interactions of CR with the atmosphere, are not included in this simulation. The energy at which the contribution of prompt muons to the sea level flux becomes equal to that of muons from $\pi , K$ decays is expected to be  $\sim 10 \ TeV$ to $\sim 10^3 \ TeV$, depending on the charm production model \cite{sine}.

In the last few years, many other CR simulation codes became available. In particular, the CORSIKA code \cite{corsika} has become a sort of standard in the cosmic ray community. However, in this context,  HEMAS has been preferred, since it was deeply used and cross-checked with MACRO data. In particular, the experimental muon 
multiplicity distribution was studied in order to infer information on
the primary cosmic ray composition \cite{macro-comp}. Moreover, some
hadronic interaction features in HEMAS, first of all the transverse momentum 
behaviour of muon parent mesons at TeV energies, were strongly checked by studying the underground muon pair distance distribution
(the so-called decoherence distribution \cite{macro-deco, tesi-max}).
Many other data-Monte Carlo comparisons, including both shower development and hadronic interaction features, can be found in \cite{tesi-max}.
An extensive study, at the level of the ANTARES detector, of the muon flux obtained with HEMAS and CORSIKA simulated showers has been performed  \cite{comparison} and has shown a general good agreement. 

\subsection{The primary CR flux and composition}
The model adopted for the primary cosmic ray energy spectrum is described in \cite{horandel}. It is a phenomenological model, named the \textit{poly-gonato} model, that uses recent results from direct and indirect measurements of cosmic rays in a wide energy range, from 10 GeV to 1 EeV.
The direct observations are used to extrapolate the energy spectra of each element to high energies. Then, the sum of all individual contributions is compared to all-particle spectra from air shower measurements.
The cut-off for each element is assumed proportional to Z, starting with the proton component at 4.5 PeV.

A simplified version of the model has been used, neglecting the elements heavier than iron and grouping the remaining ones into five mass groups:
{\it i)}   protons (A = 1);
{\it ii)}   helium  (A = 4);
{\it iii)}  CNO group (A = 14);
{\it iv)}   Mg (A = 24);
{\it v)}   Fe and heavy nuclei group (A = 56).

\begin{table}[htb]
\begin{center}

\begin{tabular}{|c|c||c|c|c|c|c|c|} \hline
        &                   & \multicolumn{6}{|c|}{Energy bins (see text)}   \\ \hline
nucleus & $\theta$ angle &  I    &   II     & III     & IV & V & VI \\ \hline \hline 
p & $0^\circ -60^\circ$       &$32 \times 10^6$ &$24 \times 10^6$&$1 \times 10^6$  & 7 $\times 10^4$  & $5 \times 10^3$  &$6 \times 10^2$    \\
        & $60^\circ-85^\circ$       &$22 \times 10^6$ &$20 \times 10^6$&$2\times 10^6$  & $1.5 \times 10^5$  &  $1.4\times 10^4$  &2 $\times 10^3$\\ \hline
He      & $0^\circ -60^\circ$       &$14 \times 10^6$ &$22 \times 10^6$&$9\times 10^5$  &  $8\times 10^4$  &  $5\times 10^3$  &$6 \times 10^2$\\
        & $60^\circ-85^\circ$       &$ 9 \times 10^6$ &$17 \times 10^6$&$1.9\times 10^6$&  1.3$\times 10^5$  & $1.3\times 10^4$  & 1.7$\times 10^3$\\ \hline
CNO     & $0^\circ -60^\circ$       &$ 4 \times 10^6$ &$13 \times 10^6$&$8\times 10^5$  &$ 6 \times 10^4$   &$5\times 10^3$   & 6$\times 10^2$\\
        & $60^\circ-85^\circ$       &$ 2 \times 10^6$ &$11 \times 10^6$&$1.4\times 10^6$&   $ 1\times 10^5$  &  $1.2\times 10^4$   & 1.4 $\times 10^3$ \\ \hline
Mg    & $0^\circ -60^\circ$       &                &$ 9 \times 10^6$&$7\times 10^5$  &  $6\times 10^4$    & $4 \times 10^3$    &5 $\times 10^2$  \\
        & $60^\circ-85^\circ$       &                &$ 8 \times 10^6$&$1.4\times 10^6$&   $ 1\times 10^5$   &  $9\times 10^3$    & 1.4 $\times 10^3$  \\ \hline
Fe      & $0^\circ -60^\circ$       &                &$ 5 \times 10^6$&$6\times 10^5$  &  $5\times 10^4$    & $4 \times 10^3$    & 5 $\times 10^2$  \\
        & $60^\circ-85^\circ$       &                &$ 4 \times 10^6$&$1.1\times 10^6$&   $6\times 10^4$   &   $5\times 10^3$   & 1.2 $\times 10^3$  \\ \hline
\end{tabular}
\end{center}
\caption{Number of generated interactions for each primary mass group, primary energy range and zenith angle interval.}
\label{tab:tabnew}
\end{table}

The minimum primary energy chosen for the simulation is 1 TeV/nucleon. 
For each mass group, the production of muons at the sea-level from primary CR has been subdivided into six energy ranges:
{\bf I)} 1 TeV - 20 TeV (for He and CNO groups the lower limits are 4 TeV and 14 TeV, respectively);
{\bf II)}  $20 - 200 $ TeV (for Mg the lower limit is 30 TeV and for iron 60 TeV); 
{\bf III)} $200 - 2 \times 10^3$ TeV ; 
{\bf IV)} $2 \times 10^3 - 2 \times 10^4$ TeV; 
{\bf V)} $2 \times 10^4 - 2 \times 10^5$ TeV; 
{\bf VI)} $2 \times 10^5 - 2 \times 10^6$ TeV.

Finally, the production of muons has been subdivided in two zenith angle intervals: 1) $0^\circ < \theta < 60^\circ$ and 2) $60^\circ < \theta < 85^\circ$.
The number of generated showers for each primary mass group, primary energy range and zenith angle interval is reported in tab. \ref{tab:tabnew}.
The general expression for the flux of each element is given by:
\begin{equation}
\frac{d\Phi_Z}{dE_0}(E_0) = \Phi^{0}_{Z} E^{\gamma_Z}_{0}\left[ 1+ \left(\frac{E_0}{\widehat{E_Z}}\right)^{\epsilon_c}\right]^{(\gamma_c-\gamma_Z)/\epsilon_c}
\end{equation}
where the parameters $\Phi^{0}_{Z}$, $\gamma_Z$ and $\widehat{E_Z}$ are the absolute flux at 1 TeV/nucleus, the spectral index and the cut-off energy (${\it knee}$), respectively and  depend on the considered nucleus. $\gamma_c$ and $\epsilon_c$,  which characterize the change in the spectrum at the cut-off energy $\widehat{E_Z}$, are assumed identical for all spectra. In Table \ref{tab:tab7}  the numerical values for all groups are listed. At the end of the shower simulation, HEMAS provides its own standard output files;  only events containing at least one muon at sea level are kept.

\begin{table}[h]
\begin{center}
\begin{tabular}{|c|c|c|c|}\hline
&   $\gamma_{Z}$   &    $\Phi^{0}_{Z}$        & $\widehat{E_{Z}} \: (GeV)$\\ \hline
p      &  -2.71             &  $8.73 \times 10^{-2} $ & $ 4.5 \times 10^{6}  $ \\
He     &  -2.64             &  $5.71 \times 10^{-2} $ & $  9 \times 10^{6}  $ \\
CNO    &  -2.68             &  $3.24 \times 10^{-2} $ & $ 3.06\times 10^{7}  $ \\
Mg     &  -2.67             &  $3.16 \times 10^{-2} $ & $ 6.48\times 10^{7}  $ \\
Fe   &  -2.58             &  $2.18 \times 10^{-2} $ & $ 1.17\times 10^{8}  $ \\ \hline
\multicolumn{2}{|c|}{$\gamma_{c}$} & \multicolumn{2}{c|}{$\epsilon_{c}$}  \\ \hline
\multicolumn{2}{|c|}{$-4.7$}      & \multicolumn{2}{c|}{$1.87$}  \\ \hline
\end{tabular}
\end{center}       
\caption{Parameters (from \cite{horandel}) of the flux of primary CR mass groups  used in this work.}
\label{tab:tab7}
\end{table}

\subsection{Propagation of muons through water}
\label{sec:PropToCan}
Muon propagation through water is performed using the MUSIC (MUon SImulation Code) program \cite{music}.\\
MUSIC  is a 3D muon propagation code originally written to describe muon propagation through rock. 
It uses recent and accurate cross sections of the muon interactions with matter and it has been modified in order to describe muon energy loss through water/ice. It takes into account energy losses due to bremsstrahlung, pair production, inelastic scattering and knock-on electron production, considered as stochastic processes if the fraction of the energy lost by muon is larger than $10^{-3}$. The angular deviation of muons is taken into account in the processes of multiple scattering, inelastic scattering and pair production.

\section{The multiplicity distribution of muons in bundles vs. depth and zenith angle \label{sec:flux} }

The multiplicity distribution of underground muons was experimentally studied with large statistics by the Frejus \cite{frejus} and the MACRO \cite{macro1} collaborations. The expected multiplicity distribution for a given primary mass and energy is known to be a negative binomial (NB) distribution. 
The observed distribution is a convolution of NB distributions, which can be described as a power law. Following the Frejus paper,  the function:
\begin{equation}
\Phi(m;h,\theta)= {K(h,\theta) \over m^{\nu(h,\theta)}} \quad with \quad \nu=  {\nu_1 \over (1+\Lambda \cdot m)} 
\label{eq:eq1}
\end{equation}
has been used as parametric formula for the flux of bundles with different number of muons $m$ at a given depth $h$ and zenith angle $\theta$. Here $K$, $\nu_1$ and $\Lambda$ are free parameters, depending on $h$ and $\theta$.
The phase space has been divided in 7 values of vertical depth $h$ (from 2.0 down to 5.0 km w.e. in steps of 0.5 km w.e.) and 9 values of zenith angle $\theta$ (from $0^\circ$ up to $80^\circ$ in steps of $10^\circ$). 
Histograms have been filled with all the muons (single or in a bundle) reaching a given vertical depth $h$, and within $\Delta\theta = \pm 1^\circ$ ($\pm 3^\circ$ for the last bin, due to statistics reasons) centred with respect to each reference zenith angle (i.e., the histograms for the value of $\theta=10^\circ$ have been filled using all muons with zenith angles $9^\circ\le \theta \le 11^\circ$) obtained with the full Monte Carlo simulation.
The 63 multiplicity distributions have been fitted with eq. \ref{eq:eq1}.

\vskip 0.4cm\noindent{\bf{The parameter $\Lambda$}}

The Frejus collaboration \cite{frejus} found, at a depth of $\simeq \ 4850$  $hg/cm^2$, a value of $\Lambda_{Fr} =0.66 \times 10^{-2} $ for shower multiplicities up to 25. 
A non zero value of $\Lambda$ would result in an incorrect increase of the flux for muon multiplicities larger than $m\sim 1/\Lambda$. 
All the fits performed are compatible with a value $\Lambda =0$. In this case, $\nu_1=\nu$ and only two free parameters must be determined in eq. \ref{eq:eq1}.

\vskip 0.4cm\noindent {\bf{The parameter $K$}}

Eq. \ref{eq:eq1} shows that the parameter $K$ has a direct physical interpretation: for $m=1$, it is the flux (in units $m^{-2} s^{-1} sr^{-1}$) of single muons coming from the  $\theta$ direction, at a vertical depth $h$. 
As a function of the vertical depth $h$ and of the zenith angle $\theta$, it can be described by the equation:
\begin{equation}
\Phi(m=1;h,\theta)= K(h,\theta)= K_0(h) cos\theta \cdot e^{K_1(h)\cdot sec\theta}\quad  (m^{-2} s^{-1} sr^{-1})
\label{eq:phi}
\end{equation}
At a given zenith angle, the flux decreases with depth and  two simple expressions for $K_0(h)$ and $K_1(h)$ have been found (the values of fitted constants are reported in Table \ref{tab:flux}):
\begin{equation}
K_0(h) = K_{0a}\cdot h^{K_{0b}} 
\label{eq:k0}
\end{equation}
\begin{equation}
K_1(h)= K_{1a}\cdot h + {K_{1b}}
\label{eq:k1}
\end{equation}

\vskip 0.4cm \noindent{\bf{The parameter $\nu$}}

The fraction of multiple muon flux with respect to the single muon flux depends on the parameter $\nu$, which, for a given vertical depth $h$, is a function of $sec\theta$: 
\begin{equation}
\nu(h,\theta)= \nu_{0}(h) \cdot e^{\nu_1(h) \cdot sec\theta}
\label{eq:nu}
\end{equation}
For a fixed zenith angle $\theta$, the parameter $\nu$ increases with increasing vertical depth $h$ as: 
\begin{equation}
\nu_0(h) = \nu_{0a} \cdot h^2 + \nu_{0b} \cdot h +\nu_{0c}  
\label{eq:nu0}
\end{equation}
\begin{equation}
\nu_1(h) = \nu_{1a} \cdot e^{\nu_{1b} \cdot h }  
\label{eq:nu1}
\end{equation}
As a comparison, the reported value \cite{frejus} from the Frejus collaboration at an average rock depth of $4.85\ km \:w.e.$ and averaged over all directions, is $\nu_{Fr}=4.63 \pm 0.11$. Equation \ref{eq:nu} (at the same water/ice depth) produces the following values: $\nu = 3.75, 4.4$ and $13.0$ for $\theta=0^\circ, 50^\circ$ and $80^\circ$, respectively.

\begin{table}
\begin{center}
\begin{tabular}
{|c||c|c|c||c|c|c|} \hline
Formula & Equat. & Name & Value & Equat. & Name & Value  \\

for the &        &      &  & & &\\ \hline
Flux & \ref{eq:k0}  & $K_{0a}$  & $7.20\times 10^{-3}$ & \ref{eq:nu0}  & $\nu_{0a}$ & $7.71\times 10^{-2}$ \\
(eq. \ref{eq:eq1})  & \ref{eq:k0}  & $K_{0b}$ & -1.927 &\ref{eq:nu0}  & $\nu_{0b}$ & 0.524  \\
     & \ref{eq:k1}  & $K_{1a}$ & -0.581  &\ref{eq:nu0}  & $\nu_{0c}$ & 2.068 \\
     & \ref{eq:k1}  & $K_{1b}$ & 0.034 &\ref{eq:nu1}  & $\nu_{1a}$ & 0.030 \\ 
     &         &          &        &\ref{eq:nu1}  & $\nu_{1b}$ & 0.470 \\ \hline 
\hline
\end{tabular}
\end {center}
\caption{\small Value of the 9 constants necessary to completely define the flux of bundles of any muon multiplicities $m$, according to eq. \ref{eq:eq1}. The parameter $\Lambda$ in eq. \ref{eq:eq1} is equal to zero, as discussed in the text.}
\label{tab:flux}
\end{table}

\begin{figure}[hbt]
\begin{center}
\epsfig{file=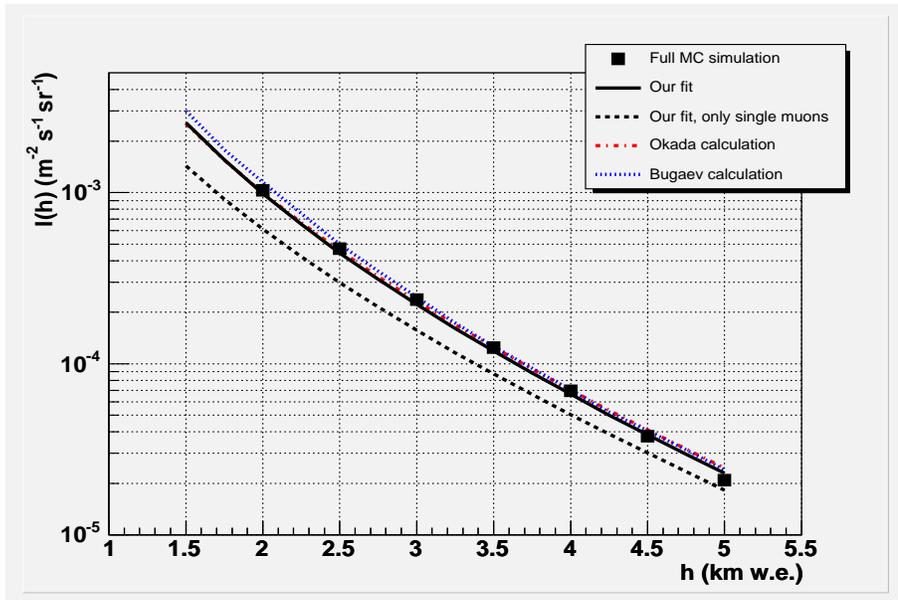,height=8cm,width=12cm}
\caption{\small Muon vertical intensity versus vertical depth. The dashed line represent the result of the  parameterisation presented in this work, eq. \ref{eq:flux}, for single muon events. The full line represents the total muon flux, which includes the contribution of bundles with any muon multiplicity $m$. Both were calculated with eq. \ref{eq:eq1} with the constants of Table \ref{tab:flux}.  
The total flux is in very close agreement with the dashed-dotted line of the Okada calculation \cite{okada} (the difference between this work and Okada is -1\% at 1.5 km w.e. and -7\% at 5.0 km w.e.). The prediction from  \cite{bugaev} is also reported as a dotted line. Both \cite{okada} and \cite{bugaev} assume a flux of single muons only. The points represent the values obtained with the full Monte Carlo simulation.  }
\label{fig:intensity}
\end{center}
\end{figure} 

\section{The depth-intensity relation \label{sec:intensity}}

\vskip 0.4cm \noindent{\bf{The vertical direction}}

\begin{figure}[hbt]
\begin{center}
\epsfig{file=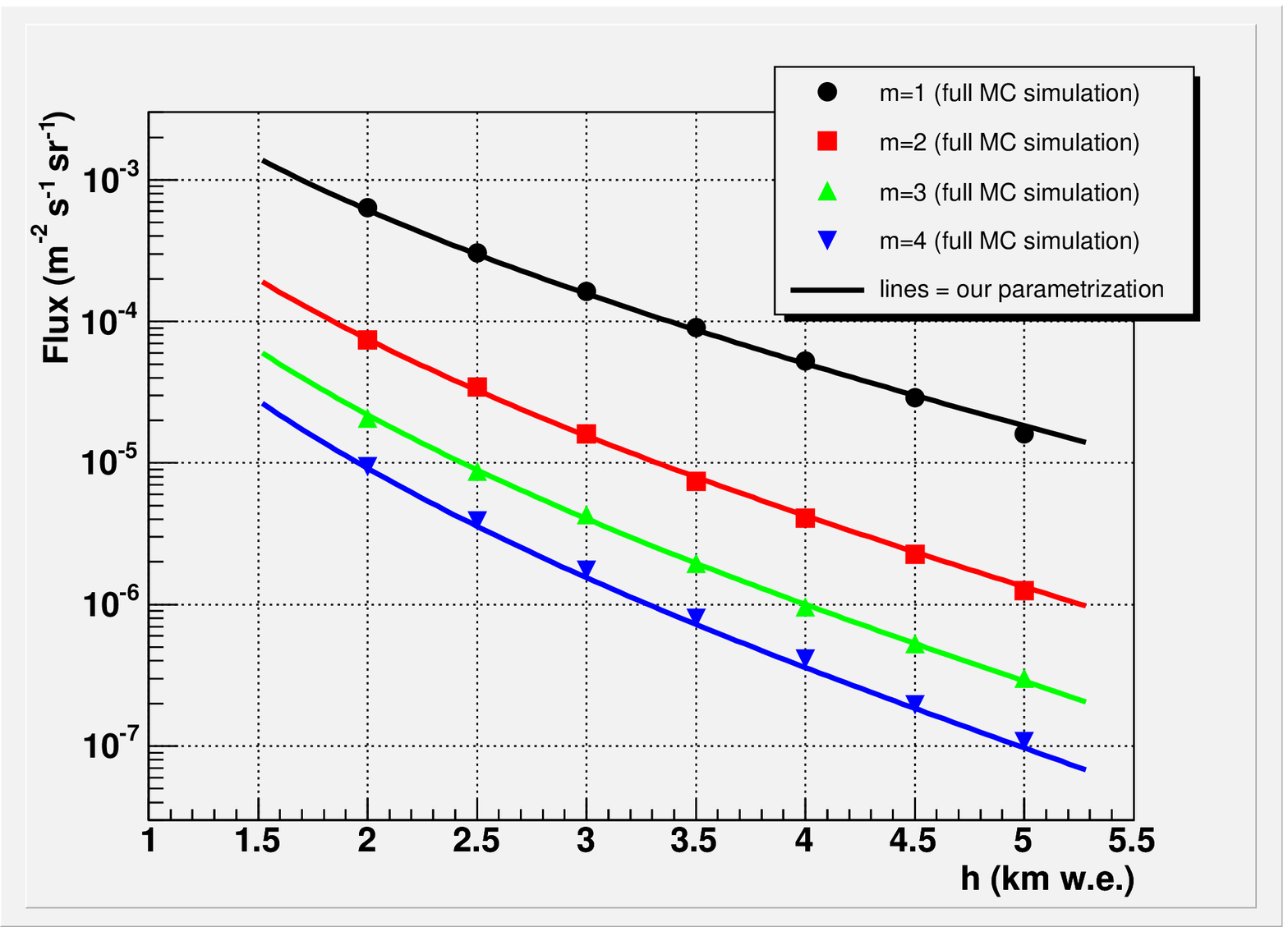,height=8cm,width=12cm}
\caption{\small Flux $I(m;h,0)$ of bundles from the vertical direction ($\theta=0^\circ$) for different values of the muon multiplicity ($m=1,2,3,4$) as a function of the vertical depth $h$. The points represent the values obtained with the full Monte Carlo (the errors are inside the symbols). The full lines were computed using eq. \ref{eq:eq1} with the constant of Table \ref{tab:flux}. }
\label{fig:multi}
\end{center}
\end{figure}

The depth-intensity relation for the vertical direction can be obtained from eq. \ref{eq:eq1}. In literature this function is called $I(h,0)$ and, according to the notation used here,  can be written as: 
\begin{equation}
I(h,0) = {\sum m \cdot \Phi(m;h,0) } = K(h,0)\cdot {\sum_m (1/m^{(\nu-1)}) \label{eq:flux} }
\end{equation}
The series at the end of eq. \ref{eq:flux} can be numerically calculated; the numerical value corresponds to the fraction of muons arriving in bundles with respect to the single ones. Fig. \ref{fig:intensity} shows the $I(h,0)$ as obtained with this  parameterisation, compared with \cite{okada,bugaev}, where all showers events are assumed as single muons only. The line representing the Okada flux is almost indistinguishable from (\ref{eq:flux}).
The seven points corresponding to the values obtained with the full Monte Carlo simulation for $\theta=0^\circ$ at different depths are also drawn. Fig. \ref{fig:multi} shows the vertical flux of bundles $I(m;h,0)$ for different values of the muon multiplicity. The ratio between the number of bundles with multiple muons  with respect to single muon events is $\sim$ 20\% at a vertical depth of 2.0 km w.e and becomes $\sim$ 11\% at depths larger than 4.0 km w.e. 

\vskip 0.4cm \noindent{\bf{ Zenith dependence }}

\begin{figure}[hbt]
\begin{center}
\epsfig{file=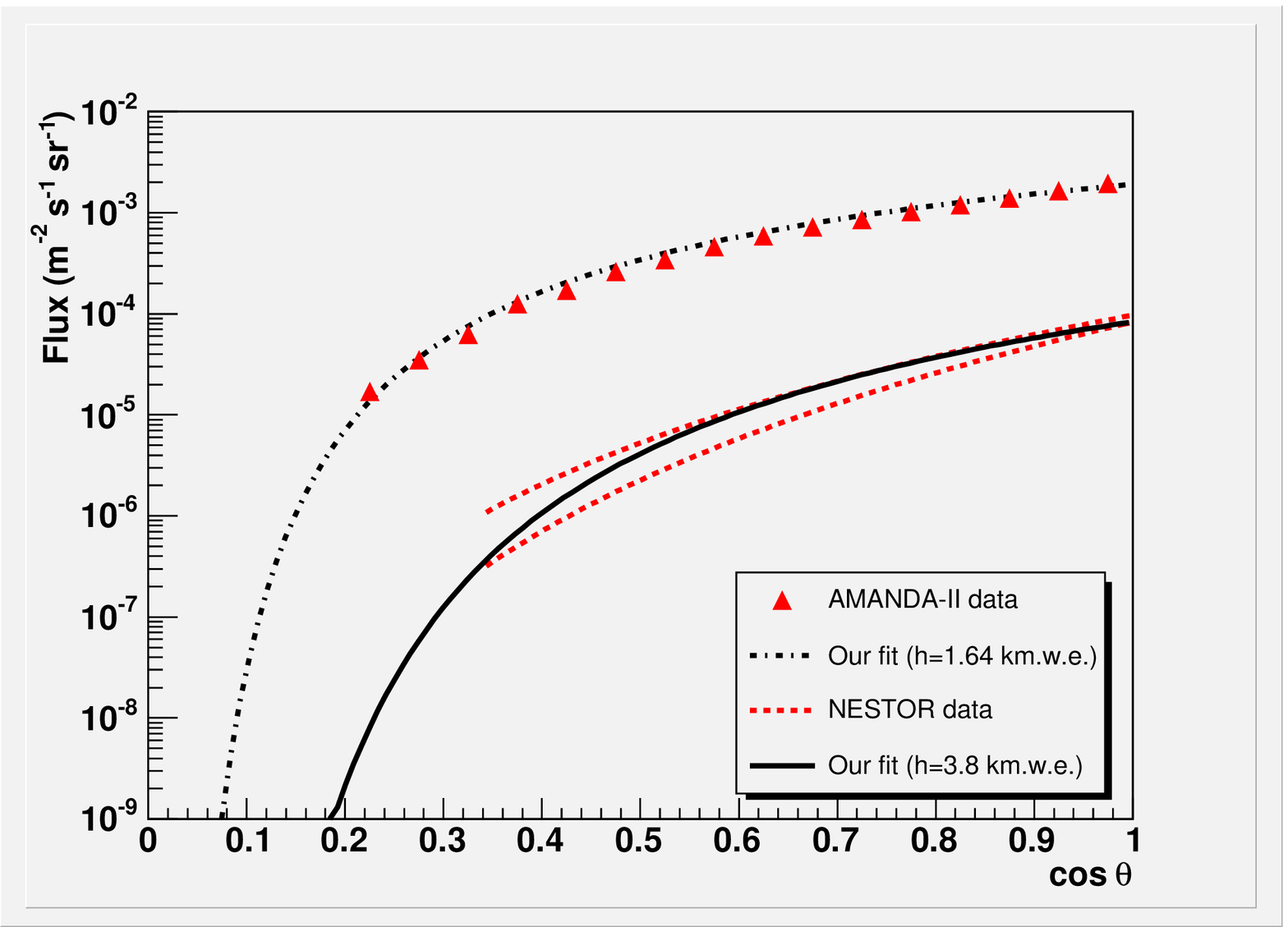,height=8cm,width=12cm}
\caption{\small Dot-dashed line: predicted total muon flux as a function of the cosine of the zenith angle $\theta$ at the depth of $h=1.64\ km \:w.e.$ The data points represent the AMANDA-II measurement \cite{amanda}; the error line is inside the marker point. Full line: predicted flux at the depth of $h=3.8\ km \:w.e.$ The dashed lines correspond to the 1-sigma error on the fit parameters of the  angular distribution as measured \cite{nestor} by the NESTOR underwater module.}
\label{fig:nestor}
\end{center}
\end{figure}

Fig. \ref{fig:nestor} shows a comparison of the predicted muon zenith angle distribution at two different depths with the measurement of the AMANDA-II under-ice experiment \cite{amanda} and of a module of the NESTOR underwater neutrino telescope \cite{nestor}. The upper dot-dashed line corresponds to the  calculation presented here, at a vertical depth $h=1.64\  km \:w.e.$ The AMANDA data (marker points) superimposed to the line have been converted to intensities relative to the underwater depths, accounting for the lower ice density ($\rho_{ice}= 0.917\ g cm^{-2}$). 

One module of the NESTOR neutrino telescope was recently deployed at a depth of $3.8\ km \:w.e.$ In \cite{nestor} a measurement of the flux of cosmic ray muons is reported as a function of the zenith angle $\theta$ according to $I(\theta)= I_o cos^\alpha (\theta)$, where $I_o=(9.0\pm 0.7) \times 10^{-5} \ (m^{-2} s^{-1} sr^{-1})$ and $\alpha=4.7\pm 0.5$. In fig. \ref{fig:nestor}, the full line represents this  calculation for $h=3.8\ km \:w.e.$, while the dashed lines indicate the 1-sigma error band on the fit parameters of the measured NESTOR angular distribution.

\section{Energy spectrum of single muons\label{sec:spectrum} }

The energy-loss processes for TeV muons are theoretically well understood and they are usually expressed as the sum of a contribution due to continuous process ($\alpha$) and a term due to catastrophic energy losses ($\beta$): 
\begin{equation}
-\langle{dE(E_\mu) \over dX}\rangle = \alpha + \beta E_\mu 
\label{eq:eloss} 
\end{equation}
Both the quantities $\alpha$ and $\beta$ depend on the medium and on $ E_\mu $.  
The expected energy distribution of underground muons, assuming a power-law for the primary beam energy, at a depth $X= h/cos\theta$ is \cite{lista}:
\begin{equation}
{ dN \over d (log_{10}E_\mu) } = G\cdot E_\mu e^{\beta X (1-\gamma)} [E_\mu + \epsilon (1-e^{-\beta X})]^{- \gamma}    
\label{eq:spectrum} 
\end{equation}
where $\gamma$ is the spectral index of the primary beam and $\epsilon= \alpha /\beta$. It must be emphasized that in this context the quantities $\epsilon, \beta$ and $\gamma$ are considered as simple parameters without a specific physical meaning.
Eq. \ref{eq:spectrum} provides the shape of the energy spectrum, where the constant $G$ represents a normalization factor:
\begin{equation}
G(\gamma,\beta,\epsilon) = 2.30\cdot (\gamma -1) \cdot \epsilon^{(\gamma-1)} \cdot e^{(\gamma-1)\cdot \beta\cdot X}\cdot (1- e^{-\beta\cdot X}  )^{(\gamma-1)} \label{eq:gi} 
\end{equation}
The overall flux of single muons is a multiplicative factor, to be calculated with eq. \ref{eq:eq1}.  
The value of the three parameters $\gamma,\beta,\epsilon$ depends both on the vertical depth $h$ and on the zenith angle $\theta$. 

As before,  the phase space has been divided in 7 values of vertical depth $h$ (from 2.0 down to 5.0 km w.e.) and 9 values of zenith angle $\theta$ (from $0^\circ$ up to $80^\circ$).  63 histograms have been filled with the energy of single muons; the histograms have been normalized to unit area, and fitted with eq. \ref{eq:spectrum}.
In order to simplify the  parameterisation,  the value of the $\beta$ parameter is fixed to $\beta= 0.420 \ (km \:w.e.)^ {-1}= 4.2\times 10^{-4}\ (hg\ cm^{-2})^{-1}$. The units of $\epsilon$ are $TeV$, and $\gamma$ is dimensionless.  The resulting muon energy is in $TeV$.

\begin{figure}[hbt]
\begin{center}
\epsfig{file=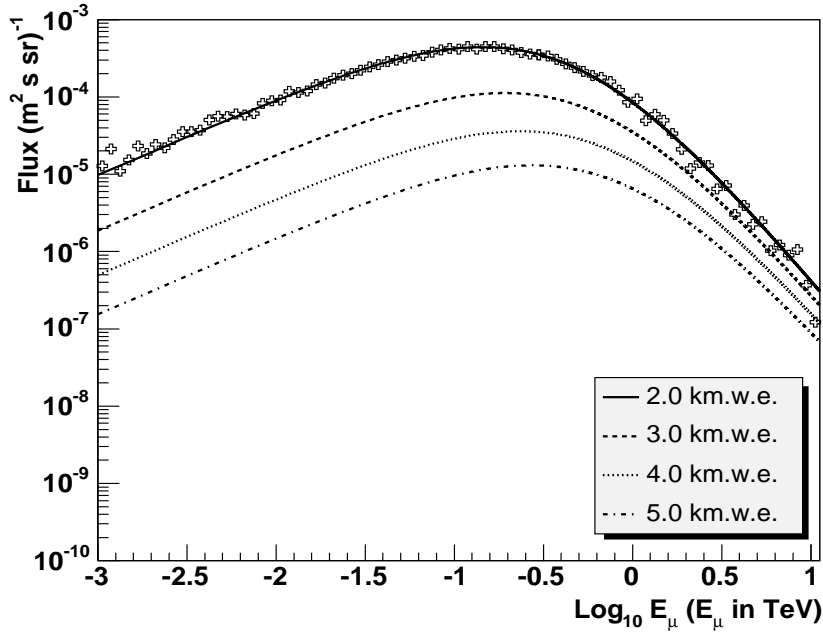,height=9cm,width=12cm}
\caption{\small Differential energy spectra of single muons from the vertical direction at various depths $(h=2,3,4$ and $5\ km \:w.e.$). The marker points (superimposed to the $h=2\ km \:w.e.$ line) correspond to the values obtained with the full Monte Carlo simulation. The lines were computed with eq. \ref{eq:spectrum} (constant of Table \ref{tab:spectrum1}), which gives the normalized shape of the distributions. The flux is obtained using a multiplicative factor computed at each different depth with eq. \ref{eq:eq1}.}
\label{fig:spectrum1}
\end{center}
\end{figure}

\vskip 0.4cm\noindent{\bf{The parameter $\gamma$}}

The parameter $\gamma$ depends on the vertical depth $h\ (km\ w.e.)$ only: its value is about 3.8 at 2.0 km w.e., and decreases to about 3.6 at 5.0 km w.e.
 $\gamma$ has been  parameterised as (see Table \ref{tab:spectrum1}): 
\begin{equation}
\gamma = \gamma(h) = \gamma_0 \cdot ln(h) + \gamma_1  \label{eq:gamma} 
\end{equation}

\vskip 0.4cm\noindent{\bf{The parameter $\epsilon$}}

The parameter $\epsilon$ depends on both the zenith angle and the vertical depth. At a given depth $h$ it shows a linear dependence on $sec\theta$, and it can be  parameterised as:
\begin{equation}
\epsilon =\epsilon(h,\theta) = \epsilon_0(h) \cdot sec\theta + \epsilon_1(h)  \label{eq:epsilon} 
\end{equation}

At a given zenith angle, $\epsilon$ increases with increasing depth $h$ as:
\begin{equation}
\epsilon_0(h) =\epsilon_{0a} \cdot e^{\epsilon_{0b}\cdot h} \label{eq:epsilon0} 
\end{equation}
\begin{equation}
\epsilon_1(h) = \epsilon_{1a} \cdot h + \epsilon_{1b}  \label{eq:epsilon1} 
\end{equation}

\begin{table}
\begin{center}
\begin{tabular}
{|c||c|c|c||c|c|c|} \hline
Formula & Equat. & Name & Value & Equat. & Name & Value  \\
for the &        &      &  & & &\\ \hline
Energy & \ref{eq:spectrum}& $\beta$  & 0.420 & 
         \ref{eq:epsilon0}  & $\epsilon_{0a}$ & 0.0304     \\
spectrum &\ref{eq:gamma} & $\gamma_{0}$  & -0.232
         &\ref{eq:epsilon0}  & $\epsilon_{0b}$ & 0.359     \\
(eq. \ref{eq:spectrum})  &\ref{eq:gamma} & $\gamma_{1}$  & 3.961
         &\ref{eq:epsilon1}  & $\epsilon_{1a}$ & -0.0077    \\

         &    &  & 
         &\ref{eq:epsilon1}  & $\epsilon_{1b}$ & 0.659     \\ \hline 
\hline
\end{tabular}
\end {center}
\caption{\small The value of the 7 constants necessary to define the (normalized) energy spectrum of single muons, eq. \ref{eq:spectrum}.}
\label{tab:spectrum1}
\end{table}

\vskip 0.4cm\noindent{\bf{Comparison of the  parametric formula (2) with the full Monte Carlo}}

At the end of the fitting procedure, using the seven constants (table \ref{tab:spectrum1}) it is possible to calculate the single muon energy distribution for any vertical depth $h$ and zenith angle $\theta$ in the range of validity.  The 63 distributions  obtained from the full Monte Carlo have been compared with the calculated ones. Each distribution has a maximum and two flex points. The energy $E_\mu^{max}$ corresponding to the maximum of the distribution is:
\begin{equation}
E_\mu^{max} = { {\epsilon (1-e^{-\beta X}) \over (\gamma - 1)} }
\label{eq:emax} 
\end{equation}
The values of the energy $E_\mu^{max}$ and that of the two flex points obtained with the parametric formula eq. \ref{eq:spectrum} differ from the corresponding points obtained with the Monte Carlo full simulation less than 3\%.

Fig. \ref{fig:spectrum1} shows the differential energy spectra of muons coming from the vertical direction at various depths. 
The lines represent the  parameterisation obtained with eq. \ref{eq:spectrum}; note that, as each integrated energy spectrum is normalized to 1, eq. \ref{eq:spectrum} gives only the shape of the distribution. A multiplicative factor from eq. \ref{eq:eq1} must be applied to obtain the flux. As an example, the points from the full Monte Carlo simulation are superimposed to the curve computed with the parametric formula for $h=2 \ km \:w.e.$ 

\begin{figure}[hbt]
\begin{center}
\epsfig{file=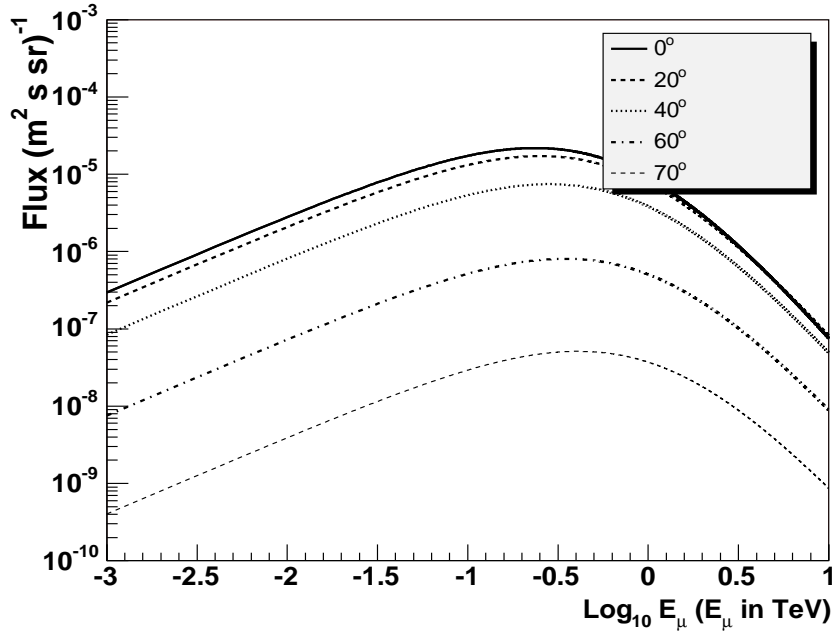,height=9cm,width=12cm}
\caption{\small Differential energy spectra of single muons at various zenith angles at a depth of $4.5\ km \:w.e. = 4500\ hg\ cm^{-2}$. 
The continuous lines have been computed with eq. \ref{eq:spectrum} and constant of Table \ref{tab:spectrum1}, which gives the normalized shape of the distributions. The flux is obtained using the multiplicative factor computed with eq. \ref{eq:eq1} at each value of the zenith angle.}
\label{fig:spectrum2}
\end{center}
\end{figure}

Fig. \ref{fig:spectrum2} shows the differential energy spectra for single muons at various zenith angles and at a depth of $4.5\ km \:w.e.$ 
The lines are obtained from the normalized  parameterisation of eq. \ref{eq:spectrum} times the absolute flux obtained with eq. \ref{eq:eq1}.

\section{Muons in bundle: the muon lateral spread \label{sec:radial} }

From theoretical and experimental considerations, it results that, in  hadron-air interactions,  particles are produced in clusters; the number of charged hadrons follows a negative binomial distribution, whose characteristics depend on the primary energy. The transverse momentum $p_t$ of the mesons follows in part 
an exponential-law distribution and in part a power-law distribution \cite{hemasdpm, macro-deco}; most of the energy is concentrated in the very forward region (i.e. near the longitudinal axis). Muons produced in the decay of secondary mesons and reaching a given depth of water follow the energy distribution of the parent mesons. As a consequence, in a muon bundle, the most energetic muons are expected to arrive closer to the axis shower. Therefore, in order to correctly parameterize the energy of muons in a bundle, the muon radial distance $R$ from the shower  axis must be taken into account.

The muon lateral distribution in a plane perpendicular to the shower axis can  be described \cite{hemas} as:
\begin{equation}
{dN \over dR} = C { R\over (R+R_0)^\alpha }
\label{eq:radial} 
\end{equation}
The average value of the radial distribution $\langle R\rangle$ depends on the parameters $R_0$ and $\alpha$: $\langle R\rangle = 2R_0/(\alpha-3$). Because of the simpler physical interpretation,  $\langle R\rangle$, instead of $R_0$, is used as fit parameter. In equation \ref{eq:radial}, C represents the normalization factor: $C=(\alpha-1)(\alpha-2) \cdot R_0^{\alpha-2}$.
The parameters $\alpha$ and $R_0$ depend on the vertical depth $h$, the zenith angle $\theta$ and on the muon multiplicity $m$ in the bundle. The $(h,\theta,m)$ phase space  has  been divided in 189 cells: the usual 7 values of vertical depth, 9 values of zenith angle and 3 multiplicities $m=2, 3$ and $>3$. 
In the following formulas  the variable $M$ is defined as: 
\begin{eqnarray}	
M = m , \quad if \ m\le 3 \nonumber \\ 
M = 4 , \quad if \ m\ge 4 
\label{eq:m} 
\end{eqnarray}	

The radial distance $R$ of each muon in a bundle of a given multiplicity $M$, reaching the seven vertical depth $h$ and in a bin of $\Delta \theta=1^\circ$ ($3^\circ$ for the last bin), with respect to one of the 9 reference values of zenith angle, was used to fill the histogram corresponding to the ($h,\theta,M$) cell.  The 189 distributions have been fitted using eq. \ref{eq:radial}.

\vskip 0.4cm\noindent{\bf{The parameter $\langle R\rangle$}}

The average value $\langle R\rangle$ of the radial distribution depends mainly on the vertical depth (it decreases when $h$ increases). Then, for a given $h$, $\langle R\rangle$ decreases with  increasing of the  muon multiplicity. Finally,  $\langle R\rangle$ does not depend on the zenith angle $\theta$ up to $\sim 50^\circ$, then it decreases with increasing $\theta$.  $\langle R\rangle$ (in units of $m$) is parameterised as:
\begin{equation}
\langle R\rangle =\rho(h,\theta,M)=\rho_0(M)\cdot h^{\rho_1}\cdot F(\theta)
\label{eq:rho} 
\end{equation}
where:
\begin{equation}
\rho_0(M) = \rho_{0a}\cdot M + \rho_{0b}
\label{eq:rho0} 
\end{equation}
\begin{equation}
F(\theta) = {1\over {e^{(\theta-\theta_0)\cdot f} +1} }
\label{eq:fteta} 
\end{equation}

\vskip 0.4cm\noindent{\bf{The parameter $\alpha$}}

The parameter $\alpha$ increases with the depth $h$ and, at a given depth, it shows a smooth decrease with increasing $M$:
\begin{equation}
\alpha = \alpha(h,M) = \alpha_{0}(M)\cdot e^{\alpha_1(M)\cdot h} 
\label{eq:alpha} 
\end{equation}
where:
\begin{equation}
\alpha_{0}(M) =   \alpha_{0a} \cdot M + \alpha_{0b} 
\label{eq:alpha0} 
\end{equation}

\begin{equation}
\alpha_1(M) = \alpha_{1a}\cdot M +\alpha_{1b} 
\label{eq:alpha1} 
\end{equation}
The value of all constants is reported in Table \ref{tab:radial}.

\begin{table}
\begin{center}
\begin{tabular}
{|c||c|c|c||c|c|c|} \hline
Formula & Equat. & Name & Value & Equat. & Name & Value  \\
for the &        &      &  & & &\\ \hline
Muon & \ref{eq:rho0}& $\rho_{0a}$  & -1.786 
     & \ref{eq:alpha0}& $\alpha_{0a}$ & -0.448    \\
lateral & \ref{eq:rho0}& $\rho_{0b}$  &  28.26
      & \ref{eq:alpha0}& $\alpha_{0b}$ &   4.969  \\
spread & \ref{eq:radial}& $\rho_{1}$  &  -1.06
     & \ref{eq:alpha1}& $\alpha_{1a}$ & 0.0194    \\
(eq. \ref{eq:radial})  & \ref{eq:fteta}& $\theta_{0}$  & 1.3  
     & \ref{eq:alpha1}& $\alpha_{1b}$ &   0.276  \\
     & \ref{eq:fteta}& f  & 10.4
     & &  &     \\
     \hline 
\hline
\end{tabular}
\end {center}
\caption{\small The value of the 9 constants necessary to define the (normalized) distribution of radial distances in bundles of muon multiplicity $M$, zenith angle $\theta$ and vertical depth $h$, eq. \ref{eq:radial}.}
\label{tab:radial}
\end{table}

\begin{figure}[htb]
\begin{center}
\epsfig{file=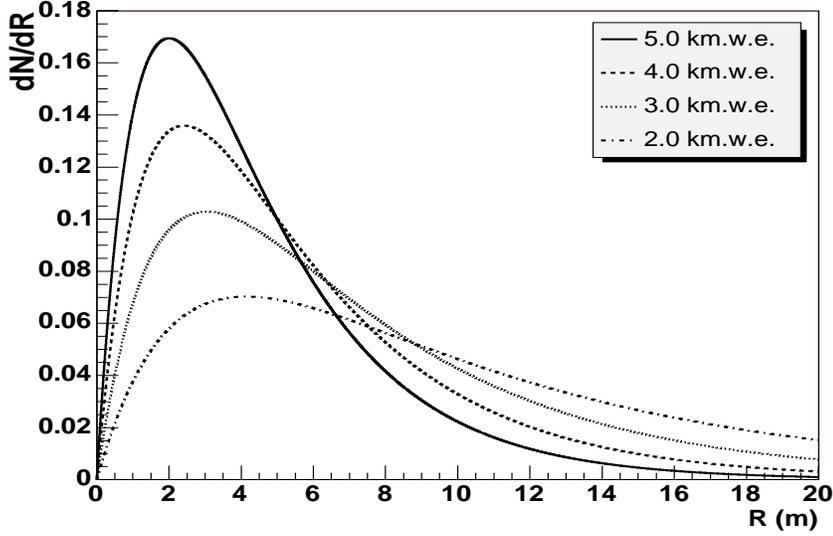,height=7.8cm,width=12cm}
\caption{\small Lateral distribution (normalized to unity area) for events with two muons in the bundle ($M=2$). The lines were computed with eq. \ref{eq:radial} and the constant of Table \ref{tab:radial} for the vertical direction ($\theta=0^\circ$) and  4 different depths: $h=2,3,4$ and $h=5\ km \:w.e.$ }
\label{fig:radial1}
\end{center}
\end{figure}

\vskip 0.2cm
Fig. \ref{fig:radial1}  shows the normalized lateral distribution of double muons for the vertical direction and at different values of the vertical depth $h$. The average value of the lateral distribution decreases when $h$ increases because the most energetic muons in the bundle arrive closer to the shower axis.

\begin{figure}[htb]
\begin{center}
\epsfig{file=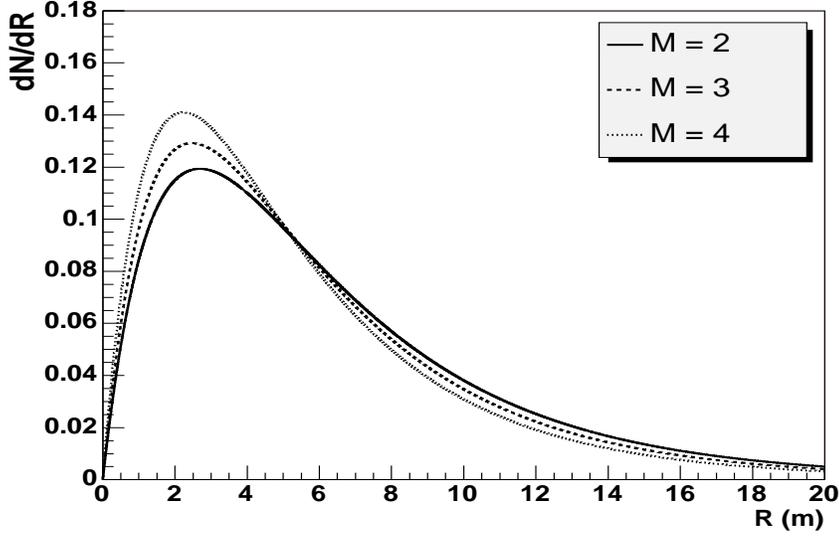,height=7.8cm,width=12cm}
\caption{\small Lateral distribution (normalized to unity area) of muons in bundles with muon multiplicities $M=2,3$ and $M=4\ (m>3)$. The lines were computed with eq. \ref{eq:radial} and the constant of Table \ref{tab:radial} for  $\theta=0^\circ$ and the vertical depth  $h=3.5\ km \:w.e.$ }
\label{fig:radial2}
\end{center}
\end{figure}

Fig. \ref{fig:radial2}  shows the normalized lateral distribution of muons with multiplicity $m= 2, 3$ and $>3 \ (M=4)$ from the vertical direction and at the depth of $3.5\ km \:w.e.$ The average value of the lateral distribution decreases when $M$ increases, because showers with large multiplicity  were originated by higher energy primary CR parents.

Note that for the  parameterisation of the energy spectrum of single muons,  the radial distance of the muon from the shower axis is not considered. However,  the formulas for the muon lateral spread (eq. \ref{eq:radial} to eq. \ref{eq:alpha1}) are valid also for the case $M=1$. 

\section{ Muons in bundle: energy spectra \label{sec:multi} }

The energy spectrum of muons arriving in bundles is described by the same function used for the single muons, eq. \ref{eq:spectrum}.
For muons in bundles, the parameters\footnote{ To avoid confusion with the symbols used for the single muon case, the three parameters are indicated with a (*)} $\gamma^*$, $\beta^*$ and $\epsilon^*$ depend, apart on the vertical depth $h$ and the zenith angle $\theta$, on the muon bundle multiplicity $M$ and on the radial distance $R$ of the muon from the shower axis. 

The $(h, \theta, M, R)$  phase space has been divided in 504 cells: 
7 values of vertical depth; 4 intervals of zenith angle ( $0^\circ-20^\circ$, $20^\circ-40^\circ$, $40^\circ-60^\circ$ and $60^\circ-80^\circ$); 3 values of muon multiplicity ($M= 2, 3, 4$); six intervals of radial distance of the muons from the shower axis: $0-5\ m$, $5-10\ m$, $10-15\ m$, $15-25\ m$, $25-45\ m$, $>45\ m$. Each $(h, \theta, M, R)$ cell correspond to one histogram which  has been filled with the value of the muon energy from the full Monte Carlo simulation. The 504 distributions have been fitted with eq. \ref{eq:spectrum}.
As in the case of single muons, $\gamma^*$, $\beta^*$ and $\epsilon^*$ have to be considered as fit parameters without any direct physical meaning.  The value $\beta^*= 0.420 \ (km \:w.e.)^{-1}= 4.2\times 10^{-4}\ (hg\ cm^{-2})^{-1}$ has been used. The unit of $\epsilon^*$ is $TeV$,  $\gamma^*$ is dimensionless.  The resulting muon energy is in $TeV$.

As for  single muons, each ${ dN \over d (log_{10}E_\mu) }$ distribution  has a value $E_\mu^{*,max}$ corresponding to the maximum (see eq. \ref{eq:emax}) and two flex points. The values of $E_\mu^{*,max}$, as a function of $\gamma^*$, and $\epsilon^*$, depend on the vertical depth $h$, on the zenith angle $\theta$, on the muon multiplicity $M$ and on the radial distance $R$ of the muon from the shower axis. In particular, keeping constant the three remaining variables, the value of $E_\mu^{*,max}$: 

\noindent - increases when the zenith angle $\theta$ increases; \par
\noindent - increases when the multiplicity $M$ increases; \par
\noindent - increases when the vertical depth $h$ increases, reaching a constant value for $h \aprge 4.5 \ km \:w.e.$; \par
\noindent - decreases when the distance $R$ of the muon with respect to the shower axis increases.

The 504 values of the $log_{10} E_\mu^{*,max}$ obtained with the parametric formula described below, differ from the corresponding values obtained with the full Monte Carlo simulation at most by 4\%. 

\begin{figure}[hbt]
\begin{center}
\epsfig{file=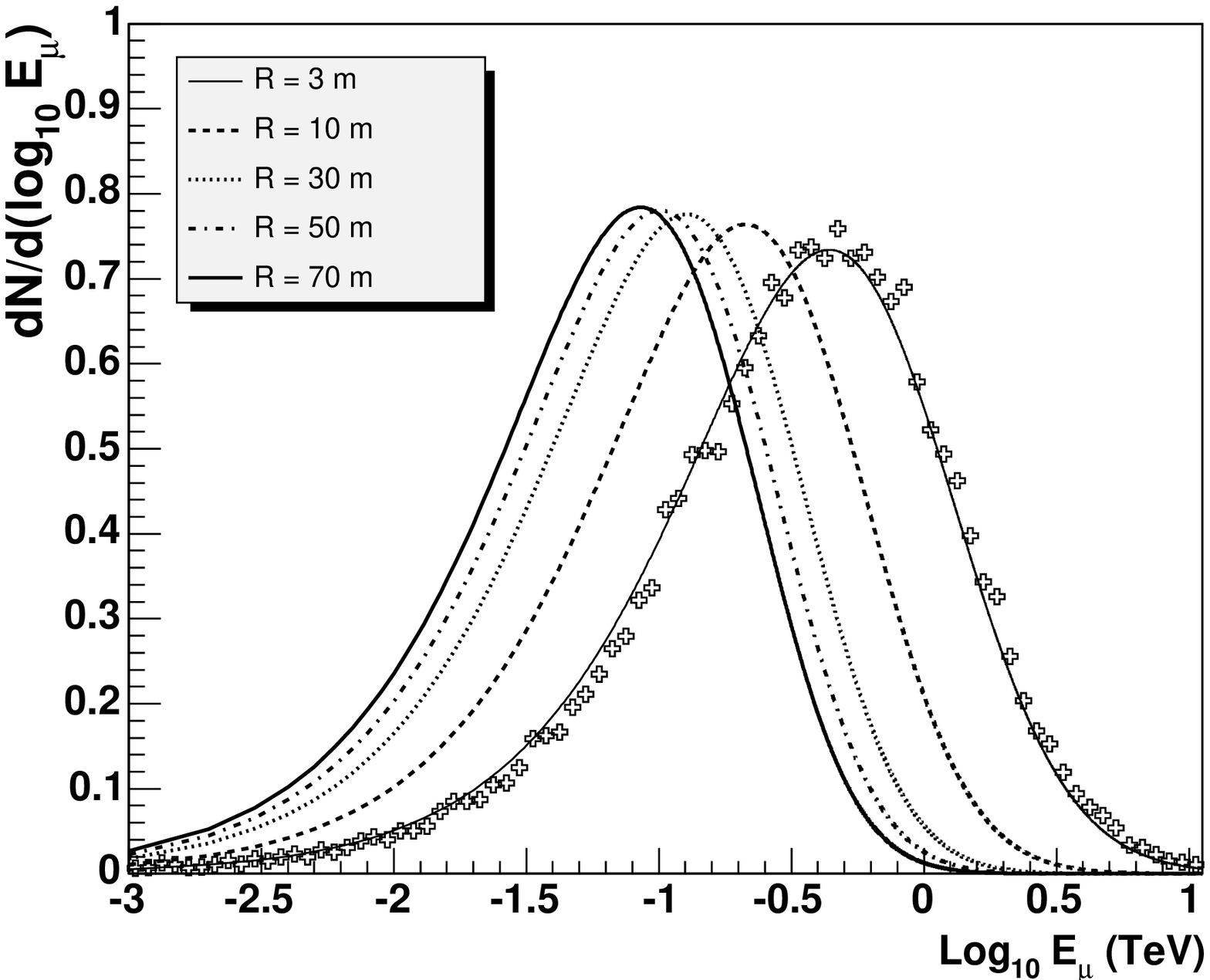,height=8cm,width=12cm}
\caption{\small Differential energy spectra of muons in bundles with multiplicity $M=2$, and five different radial distances R from the shower axis: $R=3, 10, 30, 50$ and $70 \ m$. The lines were computed with eq. \ref{eq:spectrum} and with the constants of Table \ref{tab:multimu}, assuming the vertical direction and the depth of $h=3.5\ km \:w.e. = 3500\ hg\ cm^{-2}$. The marker points superimposed for references to the line $R=3\ m$ were obtained with the full Monte Carlo simulation and used for the fitting procedure.}
\label{fig:spectrumm}
\end{center}
\end{figure}

\vskip 0.4cm\noindent{\bf{The parameter $\gamma^*$}}

As for the case of single muons, $\gamma^*$ does not depend on the zenith angle $\theta$. 
For small values of $R$ ($R \aprle 10 \ m$) it increases faster than for large values ($R \aprge 10 \ m$), where it shows a linear dependence on $R$:
\begin{equation}
\gamma^* = \gamma^*(R,h,M) = a(h) \cdot R + b(h,M)\cdot (1-{1\over 2} e^{q(h)\cdot R}) 
\label{eq:gammaa} 
\end{equation}
The slope $a$ in eq. \ref{eq:gammaa} does not depend on the muon multiplicity $M$ and it increases with the vertical depth $h$. Also the intercept $b$ increases with $h$, but with a different rate for different values of $M$.
Parameters $a$ and $b$ can be described as
\begin{equation}
a = a_o\cdot h + a_1
\label{eq:a} 
\end{equation}
\begin{equation}
b(h,M) = b_0(M) \cdot h + b_1(M)
\label{eq:b} 
\end{equation}
The dependence of $b_0$ and $b_1$ on $M$ has been  parameterised as:
\begin{equation}
b_0(M) = b_{0a} \cdot M + b_{0b}
\label{eq:b0} 
\end{equation}
\begin{equation}
b_1(M) = b_{1a} \cdot M + b_{1b}
\label{eq:b1} 
\end{equation}
The correction factor for the parameter $b$ in eq. \ref{eq:gammaa} 
has an additional parameter which depends on the depth, $q=q(h)$, as:
\begin{equation}
q(h) = q_0 \cdot h + q_1
\label{eq:q} 
\end{equation}

\vskip 0.4cm\noindent{\bf{The parameter $\epsilon^*$}}

The parameter $\epsilon^*$ does not depend on the bundle multiplicity $M$, and it grows linearly with the zenith angle $\theta$ (in radians) as:

\begin{equation}
\epsilon^* =\epsilon^*(R,h,\theta) = c(R,h) \cdot \theta + d(R,h)
\label{eq:epsilona} 
\end{equation}
For a given value of the depth $h$, both the functions $c$ and $d$ in eq. \ref{eq:epsilona} decrease with increasing value of the radial distance $R$. Thus:
\begin{equation}
c(R,h) = c_0(h) \cdot e^{c_1\cdot R}  
\label{eq:c} 
\end{equation}
\begin{equation}
d(R,h) = d_0(h) \cdot R^{d_1(h)}  
\label{eq:d} 
\end{equation}
$c_1$ is a constant, while $c_0, d_0, d_1$ depend on the vertical depth h as:
\begin{equation}
c_0(h) = c_{0a} \cdot h + c_{0b}
\label{eq:c0} 
\end{equation}
\begin{equation}
d_0(h) = d_{0a} \cdot h +  d_{0b}
\label{eq:d0} 
\end{equation}
\begin{equation}
d_1(h) = d_{1a}\cdot h + {d_{1b}} 
\label{eq:d1} 
\end{equation}
The value of all the constants necessary to define the energy spectrum of muons in bundles is reported in tab. \ref{tab:multimu}.

Fig. \ref{fig:spectrumm} shows the normalized energy distribution of double muons (M=2) for five values of the radial distance $R$ of the muons from the shower axis. A depth of $3.5 \ km \:w.e.$ and the vertical direction are assumed.
The maximum of the distribution increases when the distance of the muon from the shower axis decreases.

\begin{table}
\begin{center}
\begin{tabular}
{|c||c|c|c||c|c|c|} \hline
Formula & Equat. & Name & Value & Equat. & Name & Value  \\
for the &        &      &  & & &\\ \hline
Muon & \ref{eq:a}& $a_{0}$  &  0.0033
     & \ref{eq:c0}& $c_{0a}$ &  -0.069   \\
in bundles. & \ref{eq:a}& $a_{1}$  & 0.0079 
      & \ref{eq:c0} & $c_{0b}$ &    0.488 \\
Energy & \ref{eq:b0}& $b_{0a}$  & 0.0407  
     & \ref{eq:c}& $c_{1}$ &   -0.117 \\
spectrum & \ref{eq:b0}& $b_{0b}$ & 0.0283  
     &   &  &     \\
(eq. \ref{eq:spectrum})  & \ref{eq:b1}& $b_{1a}$  & -0.312  
     & \ref{eq:d0}& $d_{0a}$ &   -0.398  \\
     & \ref{eq:b1}& $b_{1b}$  & 6.124 
     & \ref{eq:d0}& $d_{0b}$   & 3.955    \\
     & \ref{eq:q}& $q_{0}$  &  0.0543
     & \ref{eq:d1}& $d_{1a}$   &   0.012  \\
     & \ref{eq:q}& $q_1$  &  -0.365
     & \ref{eq:d1}& $d_{1b}$   &   -0.350   \\ \hline 
\hline
\end{tabular}
\end {center}
\caption{\small The value of the 15 constants necessary to define the (normalized) distribution of muon spectrum, eq. \ref{eq:spectrum}, for bundles of muons with multiplicity $M=2,3$ and $M=4 \ (m>3)$, radial distance $R$ from the shower axis, zenith angle $\theta$ and vertical depth $h$.}
\label{tab:multimu}
\end{table}

\section{A comparison of the single and double muon energy spectrum with experimental data \label{sec:macrotrd} }

Few measurements of the average energy of underground muons are available (see \cite{macrotrd1} and references therein). The MACRO collaboration \cite{macrotrd2} used a transition radiation detector to measure the energy of single and double muons reaching the detector, crossing different depths of rock. Measurements  performed at different depths of rock accessible to the detector at different ranges of zenith angles, due to the shape of the Gran Sasso mountain coverage, show a higher average energy of muons in bundles with M=2 than for single muons.  

To compare the MACRO data with the  parameterisation presented here,  the differences in energy losses of muons in standard rock and water must be taken into account. 
The coefficient for the ionization and excitation ($\alpha$ in eq. \ref{eq:eloss}) depends on the ratio $(Z/A)$, which is assumed equal to $0.5$ for standard rock and is $10/18$ for pure water. An explicit formula for the ionization and excitation term of the muon energy losses in standard rock and water can be found in sec. 4.2 of \cite{grieder}. 
The value of the ratio $({ \alpha_R \over \alpha_W })$ (where $\alpha_R$ and $\alpha_W$ are the ionization-excitation coefficients for standard rock and water, respectively) is 0.858 (0.863) for a maximum transferable energy of 1 (100) TeV. 
The coefficient for radiative processes (bremsstrahlung, pair production and muon hadroproduction, $\beta$ in eq. \ref{eq:eloss}),  is larger in standard rock than in water. In the  1-100 TeV muon energy range, $\beta_W = (3.15\div 3.8) \times 10^{-4} \ (hg^{-1} cm^2)$ and  $\beta_R = (4.0\div 4.7) \times 10^{-4} \ (hg^{-1} cm^2)$, where $\beta_R$ and $\beta_W$ are the averaged values of the coefficient $\beta$ in standard rock \cite{lista} and water \cite{okada}, respectively.

As a  consequence, the residual energy of muons crossing a given amount of water (in $hg/cm^2$) is higher than the residual energy of muons crossing the same amount of standard rock. 
The value of the average energy $\overline E_{\mu,Rock}$  of single and double muons crossing a given depth $h$ of rock is estimated\footnote{ The formula for the average value for the energy distribution of eq. \ref{eq:spectrum} has the form of eq. \ref{eq:emax}, with the replacement of $(\gamma - 1)$ with $(\gamma - 2)$ at the denominator.}  from the corresponding value of the average energy $\overline E_{\mu,W}$ of single and double muons crossing the same depth of water, using the correction factor:
\begin{equation}
\overline E_{\mu,Rock} = \left( {\beta_W \over \beta_R}\right) \left( {\alpha_R \over \alpha_W}\right) 
\cdot  {(1-e^{-\beta_{R} h})\over (1-e^{-\beta_{W} h})} \cdot \overline E_{\mu,W} = \Gamma^W_R(h) \cdot \overline E_{\mu,W}
\label{eq:Gamma} 
\end{equation}

In Table \ref{tab:trd} the values of the average energies reported by the MACRO collaboration at four different depths are compared with the values obtained with the present  parameterisation, after the correction for the different media, eq. \ref{eq:Gamma}.
For the comparison, a second effect must be taken into account. Due to the shape of the Gran Sasso mountain, a given range of rock depth in MACRO corresponds to a range of zenith angles. In fact, for increasing zenith angles, also the average rock depth crossed by the muon increases. Using the Gran Sasso rock map \cite{macroih} (rock thickness as a function of zenith and azimuth angles),  the average value of the zenith angles for muons in each rock depth bin (column 2) is estimated, taking into account that only muons with $\theta \le 45^\circ$ were accepted for the MACRO measurement.
For double muons only,  the weighted value of the average radial distance of the muons from the shower axis, reported in column 3 has been used. 
The computed values of the correction factor  $\Gamma^W_R(h)$ are reported in column 4;  the central value of the ratios computed at 1 and 100 TeV is considered.

\begin{table}
\begin{center}
\begin{tabular}
{|c|c|c|c||c|c||c|c|} \hline
& & & & \multicolumn{2}{c||}{Single muons}& \multicolumn{2}{c|}{Double muons} \\ \hline
$h $ & $\overline \theta$  & $\overline R$ & $\Gamma^W_R $ & 
$\overline E^{MACRO}_{1\mu,Rock}$ & $\Gamma^W_R \cdot \overline E_{1\mu,W}$ & $\overline E^{MACRO}_{2\mu,Rock}$ & $\Gamma^W_R \cdot \overline E_{2\mu,W}$ \\
$(hg\ cm^{-2})$  & (deg) & $(m)$ & & $(GeV)$ & $(GeV)$& $(GeV)$& $(GeV)$\\ \hline
3280 & 3 & 5.2 & $0.768\pm 0.02$ & $250\pm 17$ & 248 & $321\pm 23$ & 329 \\
3420 & 10& 4.7 & $0.765\pm 0.02$& $262\pm 18$ & 254 & $366\pm 24$ & 350 \\
3600 & 20& 4.3 & $0.762\pm 0.02$& $278\pm 19$ & 265 & $400\pm 25$ & 383 \\
3800 & 30& 4.0 & $0.758\pm 0.02$& $283\pm 19$ & 278 & $417\pm 25$ & 412 \\ \hline 
\hline
\end{tabular}
\end {center}
\caption{\small Average energy of single (column 5) and double (column 7) muons measured by the MACRO collaboration \cite{macrotrd2} for different values of the slant depth (column 1) and of the average muon zenith angle direction (column 2). The predictions of this work   are reported in column 6 and 8 for $m=1$ and $m=2$, respectively. The calculated values were corrected to take into account the muon energy losses differences in water and rock, computed with eq.  \ref{eq:Gamma}, and reported in column 4. For double muons, the computed average distance of muons from the shower axis, reported in column 3, has been assumed .}
\label{tab:trd}
\end{table}

The excellent agreement between data and results from the parameterisation presented here (see table \ref{tab:trd}) can be partially biased by the fact that some HEMAS features were tuned using MACRO data (although data reported in \cite{macrotrd2} were not considered).

\section{Discussion and  conclusions\label{sec:conclusions} }

The knowledge of the multiple muon flux is an important requirement for neutrino telescopes. The measurement of the muon flux in a homogeneous medium, such as sea water or ice, is also of fundamental importance for the study of the muon energy losses and to test the flux of primary CR and their interactions in the upper atmosphere using the depth-intensity relation. Data obtained in deep mines or underground laboratories cannot be used to operate a fine tuning of the theoretical formulas, because of the uncertainties in rock composition and density along the muon trajectories, as pointed out in \cite{grieder}. 
Finally, the measurement of the muon flux is important to constraint the flux of high energy atmospheric neutrinos. The absolute normalization of the atmospheric neutrino flux is still affected by $\sim 30\%$ error in the 100 GeV-1 TeV range. For higher energies, the uncertainty can be even larger \cite{macronu}. 

To give an estimate of the overall uncertainty on the absolute muon flux,  formulas presented in this work are compared with experimental data and with other calculations. 
An overview of the experimental measurements of the muon vertical intensities in the deep sea is in \cite{bugaev}, where the data are also compared with a parameterisation described there (the comparison with the AMANDA-II data is in \cite{amanda}). The same data set was compared in \cite{grieder} with the  Miyake formula \cite{miyake2}, whose predictions are close to the  parameterisation of Okada \cite{okada} (shown in Fig. \ref{fig:intensity}, together with the  parameterisation discussed in the present work and the Bugaev one). The results presented here differ at most by 7\% from the Okada parameterisation, and by 15\% from the Bugaev one. The Okada and Bugaev predictions differ at most by 15\%. 

The Monte Carlo code used for simulation (HEMAS) was optimized to reproduce the multiplicity distribution of muons in bundles and the lateral distribution of muons inside the bundle at different depth of standard rock. The use of parametric formulas well reproduce all the Monte Carlo distributions. 
The average muon multiplicity and the average value of the radial distribution obtained with the full Monte Carlo differ from the results from parametric formulas less than  5\%. 
The comparison between the average muon multiplicity at different depths using HEMAS and CORSIKA agrees at the level of 10\%.

The uncertainty on the energy spectrum and on the average energy of single and multiple muons is more difficult to evaluate. The main uncertainties still arise from the hadronic interaction model and from the muon propagation through water. 
In the distributions resulting from this  parameterisation the position of some fiducial points (maximum of the distribution, flex points) differ at most by 4\% from the results obtained with the full Monte Carlo. These numerical approximation can be considered as negligible with respect to the overall theoretical uncertainties affecting the full Monte Carlo simulation.
To test the agreement between the predictions obtained with these parametric formulas and experimental data, the average energy of single and double muons measured by the MACRO collaboration at different depths \cite{macrotrd2} have been used. A correction procedure has been used to take into account the different media (water in this work, standard rock for the MACRO data). All the values agree at level of 5\%.

Finally, it should be noted that the contribution from the so called prompt muons is neglected in  HEMAS simulation. An unknown uncertainty factor due to this process should be included for muon residual energies higher than $\sim 10 \ TeV$.

In conclusion,  parametric formulas for the flux of single and multiple muons for the interval $1.5 \le h \le 5.0 \ km \:w.e.$ and for $\theta \le 85^\circ$ are given. The energy spectrum of single muons and multiple muons in bundles was also  parameterised, taking into account the dependence of the muon energy on the shower multiplicity and on the distance of the muon from the shower axis. 
The main results can be summarized as follows: $i)$ the distribution of the muon multiplicities in a bundle does depend on the vertical depth $h$ and zenith angle $\theta$; 
$ii)$ for a fixed zenith angle $\theta$, bundles with high multiplicity are suppressed when $h$ increases; 
$iii)$ the average distance of each muon in a bundle does depend on the bundle multiplicity $m$, and on $h$ and $\theta$; 
$iv)$at a given depth, the average distance of each muon in a bundle decreases slightly when the multiplicity $m$ increases. This can be qualitatively understood because bundles with high multiplicities are produced on average from primary parents of higher energies; 
$v)$the average energy of a muon in a bundle depends on the depth $h$, on the zenith angle $\theta$, on the bundle multiplicity $m$ and on the distance R of the muon from the shower axis. 

The parametric formulas presented in this paper can be easily implemented in a Monte Carlo generator which can be used to study the response of underwater/ice detectors to the flux of atmospheric muons.

\vskip 1cm

{\bf Acknowledgements}
This work was motivated through studies performed with the ANTARES and NEMO Collaborations. In particular, we would gratefully acknowledge the help and the discussions with J. Brunner, G. Carminati, S. Cecchini, C. Di Stefano, G. Giacomelli, E. Korolkova and V. Kudryavtsev. We also gratefully thank Dr. Paolo Desiati of the University of Wisconsin for providing us the reported AMANDA-II data.


\end{document}